\documentstyle[11pt,newpasp,twoside,epsf]{article}
\begin{document}

\title{Time-resolved photometry of cataclysmic variables}
\author{C. Papadaki$^{1,2}$, H.M.J. Boffin$^3$, J. Cuypers$^1$,
  V. Stanishev$^4$, Z.Kraicheva$^5$ \& V. Genkov$^5$}
\affil{
  $^1$  Royal Observatory of Belgium, 3 av. Circulaire, 1180 Brussels,
  Belgium
  $^2$  Vrije Universiteit Brussel, Astronomy Group, Pleinlaan 2, 1050 Brussel, Belgium 
  $^3$ European Southern Observatory, Garching bei Munchen, Germany
  $^4$ Department of Physics, Stockholm University, AlbaNova
  University Centre, Sweden
  $^5$ Institute of Astronomy, Bulgarian Academy of Sciences, Sofia, Bulgaria}

\begin{abstract}
We present time-resolved photometry of two cataclysmic variables whose
CCD photometric observations were obtained with the 1m telescope at
the South African Astronomical Observatory in October 2002 and August
2003 and with the 1m telescope at Hoher List in Germany.\\
Concerning MCT 2347-3144 we detect for the first time a period of 6.65h. For
V1193 Ori the 3.96 h periodicity
has for the first time been confirmed through time-resolved photometry.
\end{abstract} 
 
\section{MCT 2347-3144}
MCT 2347-3144 is a cataclysmic variable detected by the
Montreal-Cambridge-Tololo survey of subluminous blue stars (Lamontagne
et al. 2000). No
subsequent analysis and work had been done for this object so far.
\par
The CCD photometric unfiltered observations were taken with the 1m telescope at
SAAO during 4 consecutive nights in 2002 and 3 nights in 2003. Their
duration varied from 3 to 4.2h and from 4.5 to 8h hours in 2002 and
2003 respectively.
\par
The CCD frames were processed in a standard way for bias removal and
flat field corrections using the IRAF analysis packages. Aperture
photometry was conducted using the DAOPHOT procedures. The same
comparison stars were used for the data reduction in both runs. The
light curves of all the observing runs can be seen in Figures 1a and
1b for 2002 and 2003, respectively. The periodogram analysis on the
data taken in 2002, revealed a period of 0.277d or
6.65h with a
semi-amplitude of 0.144 mag. The corresponding phase diagram
is given in Figure 1c, while the folded residuals, computed after
subtraction of the corresponding periodicity from the folded data can
be seen in Figure 1d. The high amplitude derived
and the smooth phase diagram, led us to believe
its reality. However, the maximum duration of
one night's observing run being less than the periodicity we had
detected, still allowed some doubt. Therefore and in order to confirm
the 6.65h period, we needed another observing run of greater duration,
which we achieved in August 2003.
\par
The 2003 light curve is given in Figure 1b, where it is
evident that the mean magnitude of the cataclysmic variable differs
between the two years: the object appears to be one magnitude brighter
in 2003 compared to 2002. This could mean that in last August's
observing run it was in a high state or in a small outburst. Moreover 
MCT 2347-3144 shows a greater scatter in 2003 compared to 2002 and
no  evidence of the previous periodicity. The signal's absence in 2003
is also evident from Figure 1b where the sinusoidal fit of
the folded 2002 data is superimposed on the 2003 light curves.  
\vskip -1.4cm
\begin{figure}[htb]
\plotone{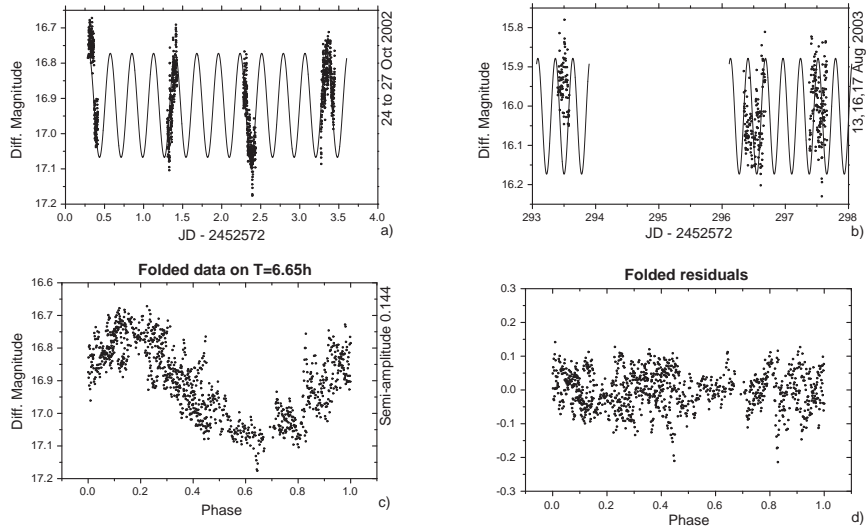}
\vskip -1.3cm
\caption{1a \& 1b: Light curves of MCT 2347-3144 with the fitted sine
  curve to the folded data of 2002, superimposed on them. 1c: Folded
  data on the periodicity of
  6.65h, the first one ever detected for the star. 1d: Folded
  residuals, coming up from the subtraction of the detected
  periodicity from the folded data in Fig. 1c.}
\end{figure}

\section{V1193 Ori}
V1193 Ori is a cataclysmic variable discovered accidentally by
M. Hamuy while making a photometric sequence in 1986 (Maza, Hamuy, \&
Ruiz 1986). The
confirmation of this CV belonging to the sub-class of nova-like
variables came one year later by Bond et al. who obtained both
high-speed photometry and spectroscopic observations. A short time
scale variability with a peak-to-peak amplitude of 0.2 mag was present
in their light curves but their power spectra showed no evidence of
any periodicities. They reported the existence of an irregular
flickering with a peak-to-peak amplitude of more than 0.15 mag. They
considered their observations insufficient for establishing the orbital
period. Warner \& Nather (1988) obtained a 3.6h photometric run
and reported rapid flickering activity of 0.25 mag but again no
orbital modulation was evident in their light curve. Later on,
time-resolved spectroscopy conducted by Ringwald, Thorstensen, \& Hamwey (1944)
revealed the existence of a 0.165d or 3.96h orbital period with an
error of 43s. 
\par
Time-resolved photometry of this object was conducted for four
consecutive nights during October 2002 at SAAO and for four more
nights at Hoher List Observatory. Their duration varied from
1.7 to 4.7 hours and from 4 to 5 hours respectively.
After the same standard procedures and Period analysis of
the 2002 light curves we were able for the first time, to photometrically
confirm the orbital period. We derive a period of 0.165d$\pm$0.00137d or
3.96h$\pm$0.0329h and a semi amplitude of 0.052 mag. The folding of the
data on this period and the corresponding residuals can be seen in
Figures 2a and 2b respectively. However, we were not able to detect
the periodicity in 2003 when the system's mean magnitude appeared to
have increased by 0.4 mag.
\vskip -1.3cm
\begin{figure}[htb]
\plotone{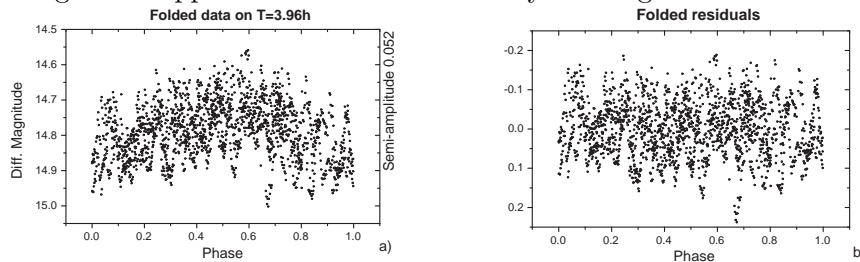}
\vskip -1.4cm
\caption{2a: Folded 2002 data on the periodicity of
  3.96h. 2b: Folded residuals, coming up from the subtraction of the detected
  periodicity from the folded data in Fig. 2a.}
\end{figure}

\par
Moreover we confirm the rapid flickering also dominating the light
curves of all photometric observations of V1193 Ori. It has a large
amplitude of 0.15 mag. From the folded residuals (Fig. 2b) it is
evident that the detection of any periodicity becomes very difficult.

\section{Discussion}
The origin of the periodicity in V1193 Ori is not clear, but by
looking at the phase diagram and taking into consideration the low
amplitude, one thought is that it could be the extra light of the side
of the secondary that is face on with the primary i.e. the irradiation
of the companion star. This speculation is strengthened by the fact
that H-$\alpha$ spectroscopy (Ringwald et al. 1994) revealed irradiation of the
secondary.

\par
Concerning MCT 2347-3144, we have checked that the absence of the
periodic signal in 2003 cannot be attributed to poorer observing
conditions. It must thus be related to the fact that the disc
brightened. In order to develop a solid interpretation
more work and observations are definitely called for.

\end{document}